\documentstyle[ijmpel,epsfig]{article}

\newcommand{\bea}{\begin{eqnarray}}
\newcommand{\eea}{\end{eqnarray}}
\newcommand{\ba}{\begin{array}}
\newcommand{\ea}{\end{array}}
\newcommand{\vx}{{\bar x}}
\newcommand{\vp}{{\bar p}}

\newcommand{\be}{\begin{equation}}
\newcommand{\ee}{\end{equation}}

\newcommand{\bi}{\bibitem}
\newcommand{\wvp}{\bar P}

\newcommand{\ms}{\mathstrut}
\newcommand{\ds}{\displaystyle}

\begin{document}
\runninghead{A quantum kinetic equation$\ldots$} 
{A quantum kinetic equation$\ldots$}
\normalsize\textlineskip
\thispagestyle{empty}
\setcounter{page}{1}

\copyrightheading{}			%{Vol. 0, No. 0 (1993) 000--000}

\vspace*{0.88truein}

\fpage{1}
\centerline{\bf A QUANTUM KINETIC EQUATION FOR PARTICLE PRODUCTION}
\centerline{\bf IN THE SCHWINGER MECHANISM}
\baselineskip=10pt
\vspace*{10pt}
\centerline{\footnotesize S. SCHMIDT\footnote{email:
basti@darss.mpg.uni-rostock.de}, D. BLASCHKE and G. R\"OPKE }
\vspace*{0.015truein}
\centerline{\footnotesize\it Fachbereich Physik, Universit\"at Rostock , D- 18051 Rostock, Germany}
\baselineskip=10pt
\vspace*{10pt}

\centerline{\footnotesize S.A. SMOLYANSKY and A.V. PROZORKEVICH}
\vspace*{0.015truein}
\centerline{\footnotesize\it  Physics Department, Saratov State University, Saratov, Russia }
\baselineskip=10pt
\vspace*{0.225truein}
\centerline{\footnotesize V.D. TONEEV\footnote{Present address: Gesellschaft 
f\"ur Schwerionenforschung, Darmstadt, Germany} }
\vspace*{0.015truein}
\centerline{\footnotesize\it Bogoliubov Laboratory of Theoretical Physics,
	Joint Institute}
\centerline{\footnotesize\it for Nuclear Research, 141980 Dubna, Russia  }
\baselineskip=10pt
\vspace*{0.225truein}

%\publisher{}{}

\abstract{A quantum kinetic equation is derived for the
description of pair production in a time-dependent homogeneous
electric field $E(t)$. As a source term, the Schwinger mechanism
for particle creation is incorporated. Possible particle
production  due to collisions and collisional damping are
neglected.  The main result is a kinetic equation of non-Markovian
character. In the low density approximation, the source
term is reduced to the leading part of the well known Schwinger
formula for the probability of pair creation. 
We discuss the momentum and time dependence of the derived source term and 
compare with other
approaches. }

\vspace{1.5cm}

%Manuscript submitted (24/09/97)

%\newpage
%\normalsize
\section{Introduction}
Particle production in ultrarelativistic heavy-ion collisions
raises a number of challenging problems. One of these
interesting questions is how to incorporate the mechanism of
particle creation into a kinetic
theory\cite{NN,BC,KM,GKM,CEKMS,Rau,Bhal,greiner,CMM}. Within the framework of a flux tube model\cite{GKM,nussinov,agi,matsui},
a lot of promising research has been carried out during the last years. In the
scenario where a chromo-electric field is generated by a
nucleus-nucleus collision, the production of parton pairs can be
described by the Schwinger mechanism\cite{Sau,HE,Sch}.
The produced charged particles  generate a
field which, in turn, modifies  the initial electric
field and may cause plasma oscillations. The
interesting question of the back reaction of this field  has been
analyzed within a field theoretical approach\cite{CM,KESCM1,KESCM2}. The results of a simple
phenomenological consideration based on kinetic equations
and the field-theoretical treatment\cite{CEKMS,KES,CHKMPA,Eis1} agree with each other.
The source term which occurs in such a modified Boltzmann
equation was derived phenomenologically in Ref.\cite{BE}.
However, the systematic derivation of this source term in
relativistic transport theory is not yet fully  carried out.
For example, recently it was pointed out that the source term may have non-Markovian character\cite{Rau,RM} even for the case of a constant electric field.

In the present work, a kinetic equation is derived in a
consistent field theoretical treatment for the time evolution of  the pair
creation in a time-dependent and spatially homogeneous electric
field. This derivation is based on the
Bogoliubov transformation for field operators between
the asymptotic in-state and the instantaneous state. In contrast
with phenomenological approaches, but in agreement with\cite{Rau,RM}, the source term of the particle
production is of non-Markovian character. The kinetic equation
derived reproduces the Schwinger result in the low density approximation in the weak field limit. 

The paper is organized as follows. In Section 2 we derive the kinetic equation
 with a new source term that describes particle production in a strong electric field  within a field theoretical approach for both  fermions
 and bosons. In Section 3 we discuss the main properties of the result and 
give numerical estimates for the case of a constant field and a model time dependent field. We summarize our results in Section 4.

%%%%%%%%%%%%%%%%%%%%%%%%%%%%%%%%%%%%%%%%%%%%%%%%%%%%%%%%%%%%%%%%
\section{Dynamics of pair creation }
\subsection{Creation of fermion pairs}
%%%%%%%%%%%%%%%%%%%%%%%%%%%%%%%%%%%%%%%%%%%%%%%%%%%%%%%%%%%%%%%%
In this section we demonstrate the derivation of a kinetic equation with a source term for fermion-antifermion production. As an illustrative example we consider electron-positron creation, however the generalization to quark-antiquark pair creation in a chromoelectric field in Abelian approximation is straightforward.
For the description of $e^+e^-$ production in an electric field we start from the QED lagrangian
\bea\label{1}
{\cal L} = {\bar \psi} i\gamma^\mu(\partial_\mu+ieA_\mu)\psi - m{\bar
\psi}\psi-\frac{1}{4}F_{\mu\nu}F^{\mu\nu}\,,
\eea
where $F^{\mu\nu}$ is the field strength, the metric is taken as $g^{\mu\nu} = {\rm diag}(1,-1,-1,-1)$ and for the
$\gamma$-- matrices we use the conventional definition \cite{bjorkendrell}. In
the following we consider the electromagnetic field as classical and
quantize only the matter field. Then the Dirac equation reads
\bea\label{2}
(i\gamma^\mu\partial_\mu-e\gamma^\mu A_\mu-m)\psi(x)=0\,.
\eea
We use a simple  field-theoretical model  to treat charged fermions
in an external electric field charactarized by the vector potential $A_\mu=(0,0,0,A(t))$ with $A(t)=A_3(t)$.
The electric field 
\bea\label{20}
E(t)=E_3(t)=-{\dot A}\ms(t)=-dA(t)/dt 
\eea
is assumed to be time-dependent but
homogeneous in space ($E_1 = E_2=0$).
This quasi-classical electric field interacts with a spinor field $\psi$ of
fermions. We look for solutions of the Dirac equation where eigenstates
are  represented  in the form
\bea\label{10} \psi^{(\pm)}_{\vp r}(x)& =& 
\bigg[i\gamma^0\partial_0+\gamma^kp_k-e\gamma^3A(t)+m\bigg] \
\chi^{(\pm)}(\vp,t) \ R_r \ {\rm e}^{i\vp\bar x},
\eea
where $k=1,2,3$ and the superscript $(\pm)$ denotes eigenstates with the  positive and negative frequencies.  Herein $R_r$ $(r = 1,2)$ is an
eigenvector of the matrix $\gamma^0\gamma^3$
\be\label{30}
R_1 =\left(\begin{array}{r}0\\1\\0\\-1\end{array}\right)\,\,,\hspace{4cm}
R_2 =\left(\begin{array}{r}1\\0\\-1\\0\end{array}\right)\,\,\, ,
\ee
so that $R^+_r R_s = 2\delta _{rs} \ .$
The functions $\chi^{(\pm)}(\vp,t)$ are related to the
oscillator-type equation
\bea
\label{40} {\ddot
\chi}^{(\pm)}(\vp,t)&=&-\bigg(\omega^2(\vp,t)+ie{\dot
A}(t)\bigg) \ \chi^{(\pm)}(\vp,t)\,\,,
\eea
where we define  the total energy $\omega^2(\vp,t)=\varepsilon_\perp^2+P_\parallel^2(t)$,
the transverse energy $\varepsilon_\perp^2=m^2+\vp^2_\perp$ and
$P_\parallel(t)=p_\parallel-eA(t)$. The solutions $\chi^{(\pm)}(\vp,t)$ of
Eq. (\ref{40}) for positive and negative frequencies are
defined by their asymptotic behavior at $t_0=t\rightarrow -\infty$, where $A(t_0) = 0$. We obtain
\be\label{60}
\chi^{(\pm)}(\vp,t) 
\sim\exp{\big(\pm i\omega_0(\vp)\,t\big)}\,\,,
\ee
where the total energy in asymptotic limit is given as $\omega_0(\vp)=\omega_0(\vp,t_0)=\lim\limits_{t\to-\infty}\omega(\vp,t)$.
 Note that the system of the spinor functions
 (\ref{10}) is complete and orthonormalized. 
The field operators $\psi(x)$ and ${\bar \psi}(x)$ can be
decomposed in the spinor functions (\ref{10})  as follows:
\be\label{70}
\psi(x)=\sum\limits_{r,\vp}
\bigg[\psi^{(-)}_{\vp r}(x)\ b_{\vp r}(t_0) +
\psi^{(+)}_{\vp r}(x) \ d^+_{-\vp r}(t_0)
\bigg] \,\,.  \ee
The operators
$b_{\vp r}(t_0),b^+_{\vp r}(t_0)$ and $d_{\vp r}(t_0),d^+_{\vp r}(t_0)$ describe
the creation and annihilation of electrons and positrons
in the in-state $|0_{in}\!>$ at $t=t_0$, and satisfy the anti-commutation relations \cite{bjorkendrell}
\be\label{71} \{b_{\vp r}(t_0),b^+_{\vp ' r'}(t_0)\}=\{d_{\vp r}(t_0),d^+_{\vp '
r'}(t_0)\}= \delta_{rr'} \ \delta_{\vp \vp '} \, . \ee
The evolution affects the vacuum state and mixes states with
positive and negative energies resulting in non-diagonal
terms that are responsible for pair creation. The
diagonalization of the hamiltonian corresponding to 
a Dirac-particle (Eq. (\ref{2})) in the homogeneous electric field (\ref{20})
is achieved by a time-dependent Bogoliubov  transformation
\bea\label{90}\ba{lclcl} b_{\vp r}(t)
&=&\alpha_\vp (t)\ b_{\vp r}(t_0)&+&
\beta_\vp (t)\ d^+_{-\vp r}(t_0)\ ,
\\ \\ d_{\vp r}(t)&=&\alpha_{-\vp} (t) \
d_{\vp r}(t_0)&-&\beta_{-\vp}(t) \
b^+_{-\vp r}(t_0)
\ea \eea
with the condition
\be\label{110}
|\alpha_\vp(t)|^2+|\beta_\vp(t)|^2=1\,\,.  \ee
Here, the operators
$b_{\vp r}(t)$ and $d_{\vp r}(t)$
 describe the creation and annihilation of quasiparticles
 at the time $t$ with the instantaneous
vacuum $|0_t\!>$. Clearly, the operator system $b(t_0),b^+(t_0);d(t_0),d^+(t_0)$ is unitary
non-equivalent to the system $b(t), b^+(t); d(t), d^+(t)$.
The substitution of Eqs.  (\ref{90}) into
Eq. (\ref{70}) leads to the new representation of the field operators
\bea\label{120} \psi(x)&=&\sum_{r,\vp}\bigg[\Psi^{(-)}_{\vp r}(x) \ b_{\vp
r}(t) + \Psi^{(+)}_{\vp r}(x) \ d^+_{-\vp r}(t)\bigg]\,\,.  \eea The link
between the new 
$\Psi^{(\pm)}_{\vp r}(x)$ and the former 
(\ref{10}) basis functions is given by a canonical transformation 
 \bea
\label{130}\ba{lcl}
\psi^{(-)}_{\vp r}(x)&=&\alpha_\vp(t) \ \Psi^{(-)}_{\vp r}(x) -
\beta^*_\vp(t) \ \Psi^{(+)}_{\vp r}(x)\,\,,\\ \\
\psi^{(+)}_{\vp r}(x)&=&\alpha^*_\vp(t) \ \Psi^{(+)}(x)
_{\vp r} + \beta_\vp(t) \ \Psi^{(-)}_{\vp r}(x)\,\,.
\ea \eea
Therefore it is justified to assume that the functions $\Psi^{(\pm)}_
{\vp r}$ have a spin structure  similar to that of $\psi^{(+)}_{\vp r}$ in
Eq. (\ref{10}),
\be
\label{150}
\Psi^{(\pm)}_{\vp r}(x)=  
\bigg[i\gamma^0\partial_0+\gamma^kp_k-e\gamma^3A(t)
+m\bigg]\phi^{(\pm)}_{\vp}(x) \ R_r \
{\rm e}^{\pm i\Theta(t)} {\rm e}^{i\vp\vx},
\ee
where the dynamical phase is defined as
\be\label{160}
\Theta(\vp,t) = \int^t_{t_0}dt'\omega(\vp,t')\,\,.
\ee
The functions $\phi^{(\pm)}_{\vp}$ are yet unknown. The substitution
of Eq. (\ref{150}) into Eqs.  (\ref{130}) leads to
the relations
\bea\label{170}\ba{lclcl}
\chi ^{(-)}(\vp,t)&=&\alpha_\vp(t) \ \phi_\vp^{(-)}(t) \ {\rm
e}^{-i\Theta(\vp,t)}&-&\beta^*_\vp(t) \ \phi^{(+)}_\vp(t) \ {\rm
e}^{i\Theta(\vp,t)}\,\,,\\ \\
\chi^{(+)}(\vp,t)&=&\alpha^*_\vp(t) \ \phi_\vp^{(+)}(t) \ {\rm
e}^{i\Theta(\vp,t)}&+&\beta_\vp(t) \ \phi^{(-)}_\vp(t) \
{\rm e}^{-i\Theta(\vp,t)}\,\,.
\ea \eea
Now we are able to find explicit expressions for
the coefficients $\alpha_\vp(t)$ and $\beta_\vp(t)$.
Taking into account that the functions $\chi ^{(\pm)}(\vp,t)$
are defined by Eq. (\ref{40}), we introduce
additional conditions to Eqs.  (\ref{170}) according to the Lagrange method
\bea\label{190}\ba{lclcl}
 {\dot \chi}^{(-)}(\vp,t)&=&-i\omega(\vp,t)\ \bigg[\alpha_\vp(t) \
\phi_\vp^{(-)}(t) \ {\rm e}^{-i \Theta(\vp,t)}&+& \beta^*_\vp(t) \
\phi^{(+)}_\vp(t) \ {\rm e}^{i\Theta(\vp,t)}\,\,\bigg]\, \,,\\ \\{\dot
	\chi}^{(+)}(\vp,t)&=&\quad i\omega(\vp,t)\ \bigg[  \alpha^*_\vp(t) \
\phi_\vp^{(+)}(t) \ {\rm e}^{i\Theta(\vp,t)}&-& \beta_\vp(t) \
\phi^{(-)}_\vp (t) \ {\rm e}^{-i\Theta(\vp,t)}\bigg]\,\,.
\ea \eea
Differentiating these equations with respect to time, using
Eqs.  (\ref{40}) and (\ref{170}) and then  choosing  as an Ansatz
\be\label{210}
\phi^{(\pm)}_\vp(t)=\sqrt{\frac{\omega(\vp,t)\pm P_\parallel(t)}{\omega(\vp,t)}} \ ,
\ee
we obtain the following differential equations for the coefficients
%\cite{PM,P,MP,CMM}:
\bea\label{240}\ba{lcl}  {\dot
\alpha}_\vp(t)&=&\quad {\ds\frac{eE(t)\varepsilon_\perp}{2\omega^2(\vp,t)}}\
\beta^*_\vp(t) \
{\rm e}^{2i\Theta(\vp,t)}\,\,,\\ \\ {\dot
\beta}^*_\vp(t)&=&-{\ds\frac{eE(t)\varepsilon_\perp}{2\omega^2(\vp,t)}}\
\alpha_\vp(t) \ {\rm
e}^{-2i\Theta(\vp,t)}\,\,.
\ea \eea
As the result of the Bogoliubov transformation we obtained the new  coefficients of the instantaneous state at the time $t$. The relations between them read 
\bea\label{220}
\ba{lcl} \alpha_\vp(t)&=&{\ds\frac{1}{2\sqrt{\omega(\vp,t) \
(\omega(\vp,t)-P_\parallel(t))}}}\bigg(\omega(\vp,t)\ \chi^{(-)}(\vp,t)+i \ {\dot
\chi}^{(-)}(\vp,t)\bigg) \ {\rm e}^{i\Theta(\vp,t)}\,\,,\\  \\
\beta^*_\vp(t)&=&-{\ds\frac{1}{2\sqrt{\omega(\vp,t) \
(\omega(\vp,t)-P_\parallel(t))}}}\bigg(\omega(\vp,t) \ \chi^{(-)}(\vp,t)-i \ {\dot
 \chi}^{(-)}(\vp,t)\bigg) \ {\rm e}^{-i\Theta(\vp,t)}\,\,.
\ea \eea

It is convenient to introduce  new operators which absorb the dynamical phase
\bea\label{260} 
B_{\vp r}(t)&=&b_{\vp r}(t) \  e^{-i\Theta(\vp,t)}\,,\\
D_{\vp r}(t)&=& d_{\vp r}(t)\  e^ {-i\Theta(\vp,t)} 
\eea
satisfying the anti-commutation relations:
\be 
\{B_{\vp r}(t),B^+_{\vp ' r'}(t)\}=\{D_{\vp r}(t),D^+_{\vp ' r'}(t)\}=
 \delta_{rr'} \ \delta_{\vp \vp '} \, . 
\ee
It is
easy to show  that these operators satisfy the Heisenberg-like  equations
 of motion 
\bea\label{270} 
\ba{lcl} {\ds\frac{dB_{\vp r}(t)}{dt}=-\frac{e
E(t)\varepsilon_\perp}{2\omega^2(\vp,t)}} \ D^+_{-\vp r}(t) + i \ [H(t), \ B_{\vp r}(t)]\,\,,
\\ \\ {\ds\frac{dD_{\vp r}(t)}{dt}=\phantom{-}\frac{e E(t)\varepsilon_\perp}
{2\omega^2(\vp,t)}}\ B^+_{-\vp r}(t)+i \ [H(t),\ D_{\vp r}(t)] \ ,
\ea 
\eea
where $H(t)$ is the hamiltonian of the quasiparticle system
\be\label{280}
H(t)=\sum_{r,\vp} \omega(\vp,t)\bigg(B^+_{\vp r}(t) \
B_{\vp r}(t)-D_{-\vp r}(t) \ D^+_{-\vp r}(t)\bigg)\,\, .
\ee
The first term on the r.h.s. of Eqs.  (\ref{270}) is caused by
the unitary non-equivalence of the in-representation and the
quasiparticle one.

Now we explore  the evolution of the distribution
function of electrons with the momentum $\vp$ and  spin $r$ defined as
\be\label{300}
f_r(\vp,t) = <0_{in}|b^+_{\vp r}(t) \ b_{\vp r}(t)|0_{in}> =
<0_{in}|B^+_{\vp r}(t) \ B_{\vp r}(t)|0_{in}>\,\,.
\ee
According to the charge conservation the distribution functions
for electrons
and positrons are equal $f_r(\vp,t) = {\bar f}_r(\vp,t)$, where
\be \label{320}
{\bar f}_r(\vp,t) = \,<0_{in}|d^+_{-\vp r}(t) \ d_{-\vp r}(t)|0_{in}>\,=\,<
0_{in}|D^+_{-\vp r}(t) \ D_{-\vp r}(t)|0_{in}>\,\,.
\ee
The distribution functions (\ref{300}) and (\ref{320}) are normalized to
the total  number of pairs $N(t)$ of the system at a given time $t$
\be\label{330} \sum \limits_{r,\vp}f_r(\vp ,t)= \sum
\limits_{r,\vp} \bar f_r(\vp ,t)=N(t)\ . \ee

Time differentiation of Eq.  (\ref{300}) leads to the following equation
\be\label{340}
\frac{d f_r(\vp,t)}{dt}= -\frac{eE(t)\varepsilon_\perp}{\omega^2(\vp,t)} \
{\rm Re}\{\Phi_r(\vp,t)\} \,\,.
\ee
Herein, we have used  the equation of motion (\ref{270}) and evaluated
the occurring commutator. The function $\Phi_r(\vp,t)$ in
Eq. (\ref{340}) describes the creation  and
annihilation   of an electron-positron pair
in the external electric field $E(t)$ and is given as
\be \label{350}
\Phi_r(\vp,t)= <0_{in}|D_{-\vp r}(t) \
B_{\vp r}(t)|0_{in}>\,\,.\ee
It is straightforward to evaluate the derivative of this function.
Applying
the equations of motion (\ref{270}), we obtain
\be\label{360}
\frac{d\Phi_r(\vp,t)}{dt}=\frac{e
E(t)\varepsilon_\perp}{2\omega^2(\vp,t)}\bigg[2f_r(\vp,t)-1\bigg]-2i\omega(\vp,t) \
\Phi_r (\vp,t)\,\,,\ee
where because  of charge  neutrality of the system, the relation
$f_r(\vp,t) = {\bar f}_r(\vp,t)$ is used. The solution of
 Eq. (\ref{360}) may be written in the following integral form
\be\label{370}
\Phi_r(\vp,t) = \frac{\varepsilon_\perp}{2}\int_{t_0}^t dt'\frac{e
E(t')}{\omega^2(\vp,t')}\bigg[2f_r(\vp,t')-1\bigg]{\rm e}^{
2i[\Theta(\vp,t')-\Theta(\vp,t)]}\,\,.
\ee
The functions $\Theta (\vp,t)$ and $\Theta (\vp,t')$ in 
Eq. (\ref{370}) can be taken at $t_0$ (see the definition (\ref{160})).
Hence with $A(t_0)=0$, 
the function $\Phi_r(\vp ,t)$ vanishes at $t_0$.
Inserting Eq. (\ref{370}) into the r.h.s of Eq. (\ref{340})
we obtain
\be\label{380} \frac{df_r(\vp,t)}{dt}=
\frac{e
E(t)\varepsilon_\perp}{2\omega^2(\vp,t)}\int_{-\infty}^t dt' \frac{e
E(t')\varepsilon_\perp}{\omega^2(\vp,t')}\bigg[1-2f_r(\vp,t')\bigg]\cos
\bigg(2[\Theta(\vp,t)-\Theta(\vp,t')]\bigg)\,\,.
\ee
Since the distribution function obviously does not depend on spin (\ref{380}), we can
define: $f_r = f$.
With the substitution $f(\vp,t)\to F(\wvp,t)$,
where the 3-momentum is now defined as
$\wvp (p_\perp, P_\parallel(t))$,
the kinetic equation (\ref{380}) is reduced to its final form:
\be\label{400}
\frac{dF(\wvp,t)}{dt}=\frac{\partial F(\wvp,t)
}{\partial t}+eE(t)\frac{\partial F(\wvp,t) }{\partial P_\parallel(t)}=
{\cal S_{(-)}}(\wvp,t)\,\,,
\ee
with the  Schwinger source term
\be
\label{410}
{\cal S_{(-)}}(\wvp,t) = \frac{e
E(t)\varepsilon_\perp}{2\omega^2(\vp,t)}\int_{-\infty}^t dt' \frac{e
E(t')\varepsilon_\perp}{\omega^2(\vp,t')}\bigg[1-2F(\wvp,t')\bigg]\cos
\bigg(2[\Theta(\vp,t)-\Theta(\vp,t')]\bigg)\,\,.
\ee
 Recently in Ref.\cite{Rau}, a kinetic
equation similar to (\ref{400}) has been derived within a projection
operator formalism for the case of a time-{\em in}dependent
electric field  where it was first noted that this source
term has non-Markovian character. As well as in this method the multiple pair creation is not considered.
The presence of the Pauli blocking factor $[1-2F(\wvp,t)]$ in the source term
has been obtained earlier in Ref.\cite{KESCM2}.
We would like to emphasize the closed form of the kinetic equation in the
present work where the source term does not include the anomalous
distribution functions (\ref{350}) for fermion-antifermion pair creation
(annihilation).

%%%%%%%%%%%%%%%%%%%%%%%%%%%%%%%%%%%
\subsection{Creation of boson pairs}
%%%%%%%%%%%%%%%%%%%%%%%%%%%%%%%%%%%%

In this subsection we derive the kinetic equation with source term for the 
case of  pairs of charged bosons in a strong electric field. 

The Klein-Gordon equation reads
\be\label{kg}
\bigg((\partial^\mu + ieA^\mu)(\partial_\mu+ieA_\mu)+m^2\bigg)\phi(x)=0\,.
\ee
The solution of the Klein-Gordon equation in the presence of the electric
field defined by the vector potential $A_\mu =(0,0,0, A(t))$
is taken in the form \cite {bjorkendrell}
\be\label{A1}
\phi^{(\pm)}_{\vp}(x)=[2\omega(p)]^{-1/2}\ e^{i\bar
x\vp} g^{(\pm)}(\vp ,t)\ ,  
\ee 
where the functions $g^{(\pm)}(\vp ,t)$
satisfy  the oscillator-type equation with a variable frequency
\be\label{A2} 
\ddot
g^{(\pm)}(\vp ,t)+\omega^2 (\vp,t) \ g^{(\pm)}(\vp ,t)=0\ .
\ee
Solutions of Eq. (\ref{A2}) for positive and negative frequencies are defined
by their asymptotic behavior at $t_0=t\to -\infty$ similarly to
Eq. (\ref{60}). 

 The field operator in the in-state is defined as 
\be\label{A4}
\phi (x)=\int d^3p \ [\ \phi^{(-)}_{\vp}(x) \ a_{\vp}(t_0)+
\phi^{(+)}_{\vp}(x) \ b^+_{\-vp}(t_0)\ ]\,.
\ee
The diagonalization of the hamiltonian corresponding to the instantaneous stateis achieved by the transition to
the quasiparticle representation. The Bogoliubov transformation for
creation and annihilation operators of quasiparticles
has the following form
\be\label{A5}\ba{lclcl}
a_{\vp }(t)
&=&\alpha_\vp (t)\ a_{\vp }(t_0) &+&\beta_\vp (t)\ b^+_{-\vp }(t_0)\ ,
\\ \\ b_{-\vp }(t)&=&\alpha_{-\vp}(t)\ b_{\vp }(t_0)&+&
\beta_{-\vp}(t)\ a^+_{-\vp }(t_0)\  \ea \ee
with the condition \be\label{A6}
|\alpha_\vp(t)|^2-|\beta_\vp(t)|^2=1\,\,.  \ee
The derivation of the Bogoliubov coefficients $\alpha$ and $\beta$ is similar to  that given in the previous subsection.  
We obtain the equations of
motion for the coefficients of the canonical transformation (\ref{A5}) as follows
\bea\label{A12}
\dot\alpha_\vp (t)&=&\frac{\dot\omega(\vp ,t)}{2\omega(\vp ,t)} \ 
\beta^*_\vp(t) \
e^{2i\Theta (\vp ,t)}\,,\\
 \dot\beta_\vp(t)&=& \frac{\dot\omega(\vp ,t)}{2\omega(\vp ,t)}
\ \alpha^*_\vp (t) \ e^{2i\Theta (\vp ,t)}\,. 
\eea

Following the derivation for the case of fermion production, 
we arrive at the final result for the source term in the bosonic case 
\be
\label{440}
{\cal S_{(+)}}(\wvp,t) = \frac{e E(t)\varepsilon_\perp}{2\omega^2(\vp ,t)}
\int_{-\infty}^t dt' \frac{e
E(t')\varepsilon_\perp}{\omega^2(\vp ,t')} \ \bigg[1+2F(\wvp,t')\bigg]\cos
\bigg(2[\Theta(\vp ,t)-\Theta(\vp ,t')]\bigg)\,,
\ee
which differs from the fermion case just by the sign in front of the distribution function due to the different statistics of the produced particles.

%%%%%%%%%%%%%%%%%%%%%%%%%%%%%%%%%%%%%%%%%%%%%%%%%%%%%%%%%%%%%%%%%%%%%%%%%%%
\section{Discussion of the source term}
%%%%%%%%%%%%%%%%%%%%%%%%%%%%%%%%%%%%%%%%%%%%%%%%%%%%%%%%%%%%%%%%%%%%%%%%%%%%
\subsection{Properties of the source term}
%%%%%%%%%%%%%%%%%%%%%%%%%%%%%%%%%%%%%%%%%%%%%%%%%%%%%%%%%%%%%%%%%%%%%%%%55

We can combine  the results for fermions (\ref{410}) and for bosons (\ref{440}) into a single  kinetic equation 
\be\label{445}
\frac{dF_{(\pm)}(\wvp,t)}{dt}=\frac{\partial F_{(\pm)}(\wvp,t)
}{\partial t}+eE(t)\frac{\partial F_{(\pm)}(\wvp,t) }{\partial P_3}=
{\cal S_{(\pm)}}(\wvp,t)\,\,.
\ee
Here, the upper (lower) sign corresponds to the Bose-Einstein (Fermi-Dirac)
statistics. Based on microscopic dynamics, these kinetic equations are
exact within the approximation of a
time-dependent homogeneous electric field and the  neglect of collisions. The Schwinger source terms (\ref{410}) and (\ref{440})
are characterized by the following features:
\begin{enumerate}
\item{The kinetic equations (\ref{445}) are of non-Markovian type due
to the explicit dependence of the source terms on the whole pre-history
via the statistical factor $1\pm2F({\bar P},t)$ for fermions or bosons, respectively. The memory effect is expected to lead to a modification of particle pair creation as compared to the (Markovian) low-density limit, where the statistical factor is absent.}
\item{The difference of the dynamical phases,
$\Theta(\vp ,t)-\Theta(\vp ,t')$, under the integrals (\ref{410}) and 
(\ref{440}) generates high frequency oscillations.}
\item{ The appearance of such a source term leads to entropy
production due to pair creation (see also Rau\cite{Rau}) and therefore the time reversal
symmetry should be violated, but it does not result in any monotonic
entropy increase (in absence of collisions).}
\item{The source term  and the distribution functions have a non-trivial momentum dependence resulting in the fact that particles are produced not only at rest as assumed in previous studies, e.g. Ref.\cite{KES}.}
\item{In the low density limit and in the simple case of a constant electric field we reproduce the pair production rate given by
Schwinger's formula \be\label{430} {\cal S}^{{\rm cl}}=\lim_{t\to 
+\infty}(2\pi)^{-3}g\int d^3P \ {\cal S}(\wvp,t)=\frac{e^2E^2}{4\pi^3}\exp 
\bigg(-\frac{\pi m^2}{|eE|}\bigg) \ .
\ee}
\end{enumerate}

As noted above, Rau\cite{Rau} found that the source term has a non-Markovian behaviour  by deriving the production rate within a projector method. In the limit of a constant field our results agree with Ref.\cite{Rau}.  In our approach the electric field is treated as a general time dependent field and hence there is no {\em a priori} limitation to constant fields. However our result allows to explore the influence of any time-dependent electric field on the pair creation process. It is important to note that in general this time dependence should be  given by a selfconsistent solution of the coupled field equations, namely the Dirac (Klein-Gordon) equation {\em and} the Maxwell equation. This would incorporate back reactions as mentioned in the introduction.  However, the solution of such a system of equations is beyond the scope of this work. Herein we will restrict ourselves to the study of some features of the new source term.

Finally we remark that the source term is characterized by  two time scales:
the memory time 
\be
\tau_{mem}\sim \frac{\varepsilon_\perp}{eE}
\ee
and the production interval
\be
\tau_{prod}=1/<S_{(\pm)}>\,,
\ee
with $<S_{(\pm)}>$ denoting the time averaged production rate.
As long as $E<<m^2/e<\varepsilon^2_\perp/e$, the particle creation process is Markovian: $\tau_{mem}<<\tau_{prod}$. This results for constant fields agree with those of Rau\cite{Rau}.
\begin{figure}[hbtp]
\centering{\epsfig{figure=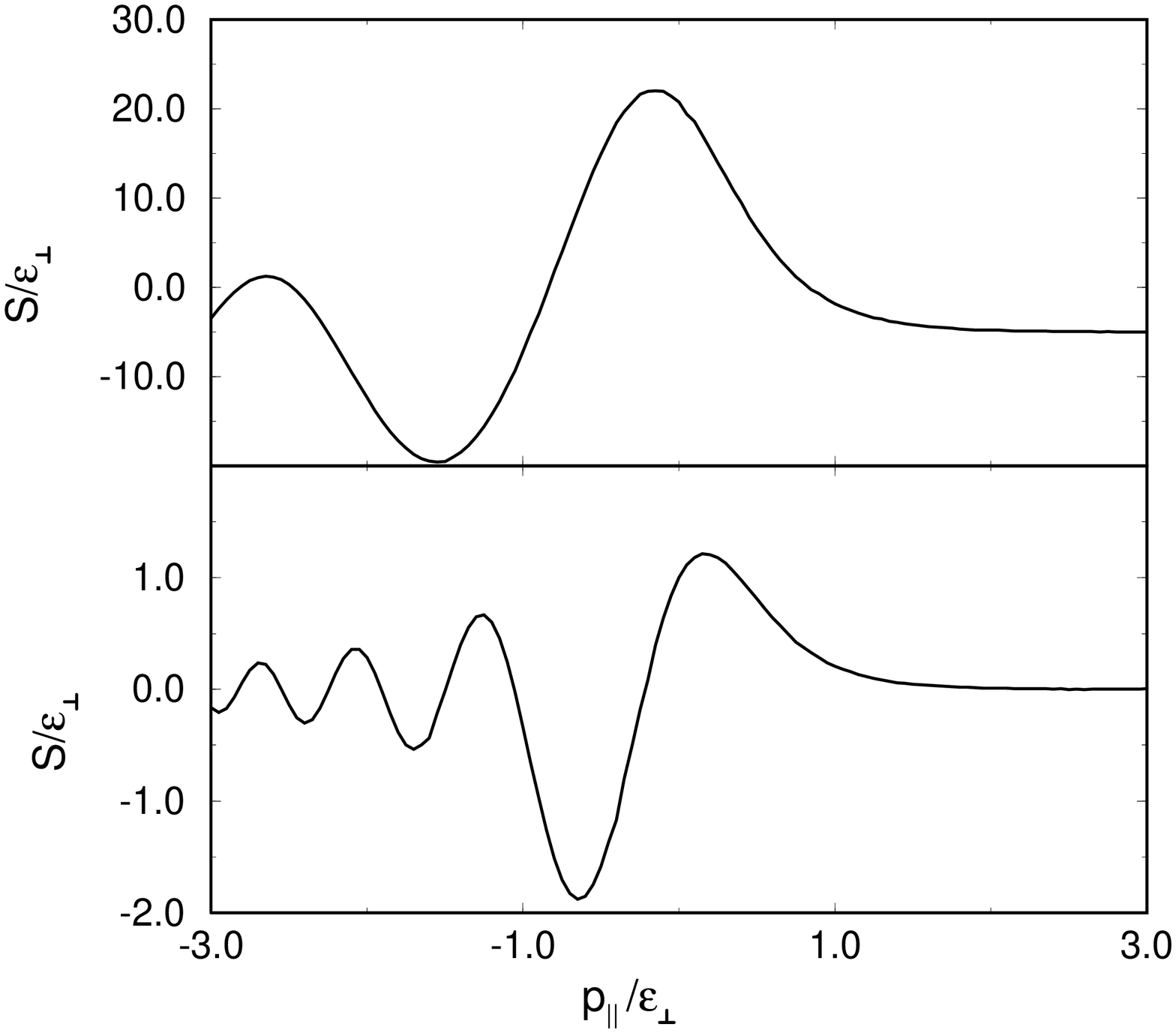,height=7.0cm,width=7.0cm,angle=-90}}

\vspace{0.3cm}

\fcaption{The pair production rate as a function of parallel momentum for a constant, strong field (upper plot: ${\tilde E}_0 = 1.5$) and a  weak field (lower plot: ${\tilde E}_0 = 0.5$) at ${\tilde t}=0$.\label{fig1}}
\end{figure}

%%%%%%%%%%%%%%%%%%%%%%%%%%%%%%%%%%%%%%%%%%%%%%%%%%%%%%%%%%%%%%%%%%%
\subsection{Numerical results}
%%%%%%%%%%%%%%%%%%%%%%%%%%%%%%%%%%%%%%%%%%%%%%%%%%%%%%%%%%%%%%%%%
In order to study the new source term, we consider two different cases, namely  a constant  field  and a time dependent electric field.
For the numerical evaluation we start with Eq.  (\ref{410}) assuming
a dilute system, $F = 0$, and 
\be
\label{dimsource}
{\tilde {\cal S}}({\tilde p_\parallel},{\tilde t})=\frac{S({\tilde p_\parallel},{\tilde t})}{\varepsilon_\perp}= 
\frac{ {\tilde E}({\tilde t}) }{ 2{\tilde \omega}^2({\tilde
p_\parallel},{\tilde t}) }
\int_{-\infty}^{\tilde t} d{\tilde t}' \frac{
{\tilde E}({\tilde t}')}{{\tilde \omega}^2({\tilde p_\parallel},{\tilde
t}')}\cos\bigg(2[\Theta({\tilde p_\parallel},{\tilde t})-\Theta({\tilde
p_\parallel},{\tilde t}')]\bigg)\,\,,
\ee
where we have introduced dimensionless variables
\be
{\tilde E}({\tilde t}) =
eE({\tilde t})/\varepsilon_\perp^2\,,\hspace{1cm} {\tilde t} = t\varepsilon_\perp\,,
\ee
\be
{\tilde
p_\parallel}=  p_\parallel/\varepsilon_\perp\,,\hspace{1cm}{\tilde \omega} =
\omega/\varepsilon_\perp\,.
\ee
This notation is convenient to distinguish the weak field (${\tilde E}<1$) and strong field (${\tilde E}>1$) limits.
A particularly simple result is obtained if we assume a constant field,
\be
{\tilde A}({\tilde t}) = A({\tilde t})/\varepsilon_\perp = {\tilde t}{\tilde E}_0/e\,,
\ee
where
the electric field does not depend on time and the energy is given as 
\be\label{energycon}
{\tilde
\omega}_0^2({\tilde p_\parallel},{\tilde t}) = 1 + ({\tilde p_\parallel} - {\tilde E}_0{\tilde t})^2\,.
\ee
For the source term we obtain
\be\label{sourcecon}
 {\tilde {\cal S}}({\tilde p_\parallel},{\tilde t}) = \frac{{\tilde E}_0^2}
{2{\tilde \omega}_0^2({\tilde p_\parallel},{\tilde t})}
\int_{-\infty}^{\tilde t} d{\tilde t}'\frac{1}{{\tilde
\omega}_0^2({\tilde p_\parallel},{\tilde t}')}\cos\bigg(2\int_{\tilde t'}^{\tilde
t}d{\tilde t}''{\tilde \omega}_0({\tilde p_\parallel},{\tilde t''})\bigg)\,.
\ee
\begin{figure}[hbtp]
\centering{\epsfig{figure=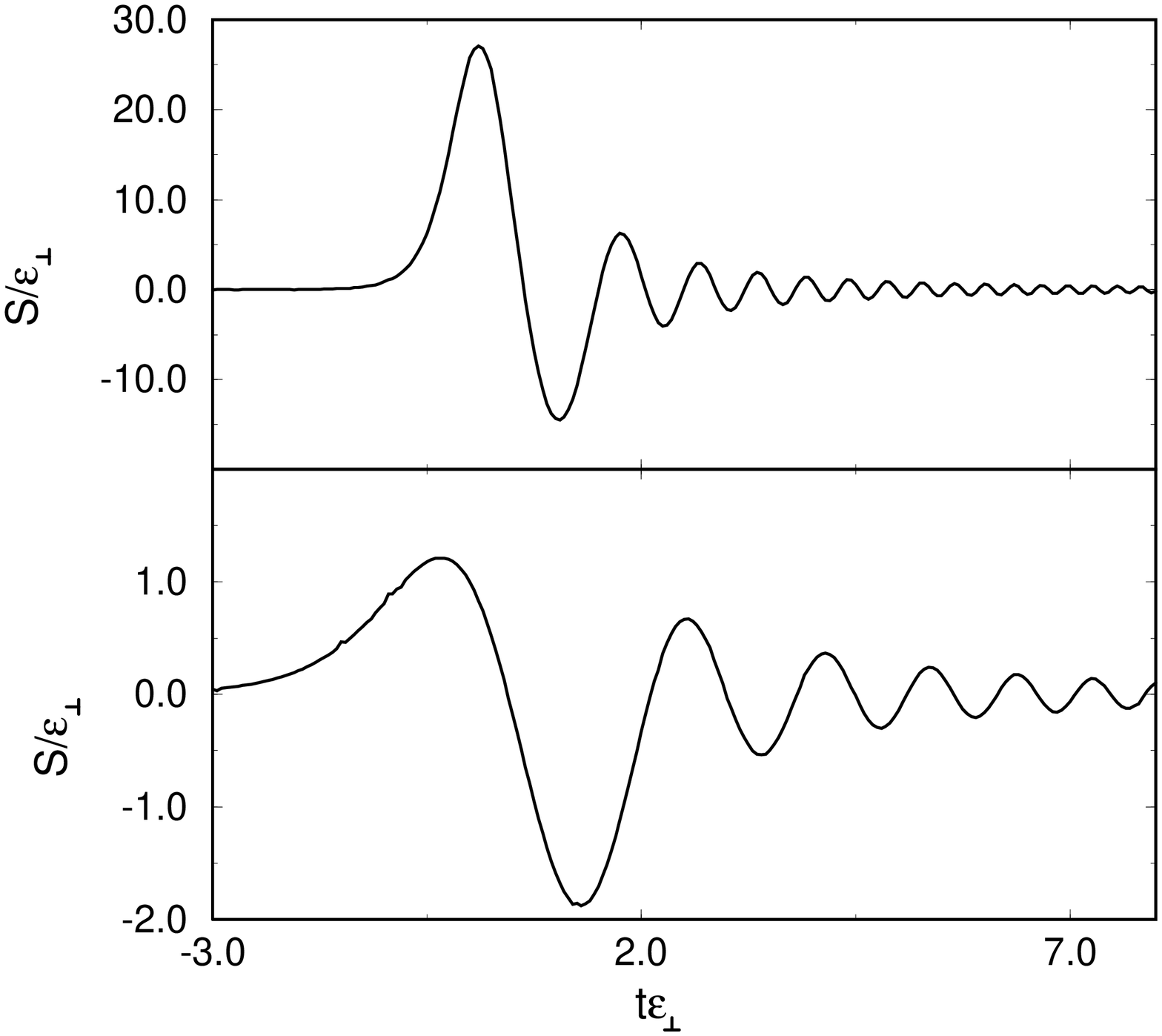,height=7.0cm,width=7.0cm,angle=-90}}

\vspace{0.5cm}

\fcaption{The pair production rate as a function of time for a constant, strong field (upper plot: ${\tilde E}_0 = 1.5$) and a  weak field (lower plot: ${\tilde E}_0 = 0.5$) at zero parallel momentum.\label{fig3}}
\end{figure}
In Fig. \ref{fig1} we plot the particle production rate as function of the parallel momentum $p_\parallel$ for a weak constant field  and a strong field, respectively. The rates are normalized to be of the order of one for ${\tilde E}_0 = 0.5$ at zero values of both momentum and time. The production rate is positive when particles are produced with positive momenta. Pairs with negative momenta are moving against the field and hence get annihilated, clearly to be seen in the negative production rate. These results mainly agree with those of Ref.\cite{Rau}, since no time dependence was considered. Note that using the prefactor and field strengths chosen by Rau, we  reproduce exactly the results given in Ref.\cite{Rau}.   

In considering  the time dependence of the source term, we go beyond the 
analysis of Ref.\cite{Rau}.  
In Fig.  
\ref{fig3}, we display the time dependence of the production rate  at zero momentum.
The maximum of the production rate is concentrated around zero and shows an
oscillating behaviour for large times. Indeed it is possible to write Eq. (\ref{sourcecon}) in terms of the Airy function because the constant field provides an analytical solution for the dynamical phase difference using the energy given in Eq. (\ref{energycon}).
 The production of pairs  
for strong fields is larger than that of weak fields, and the typical time
period of the oscillations  is smaller. 
 \begin{figure}[hbtp]
\centering{\epsfig{figure=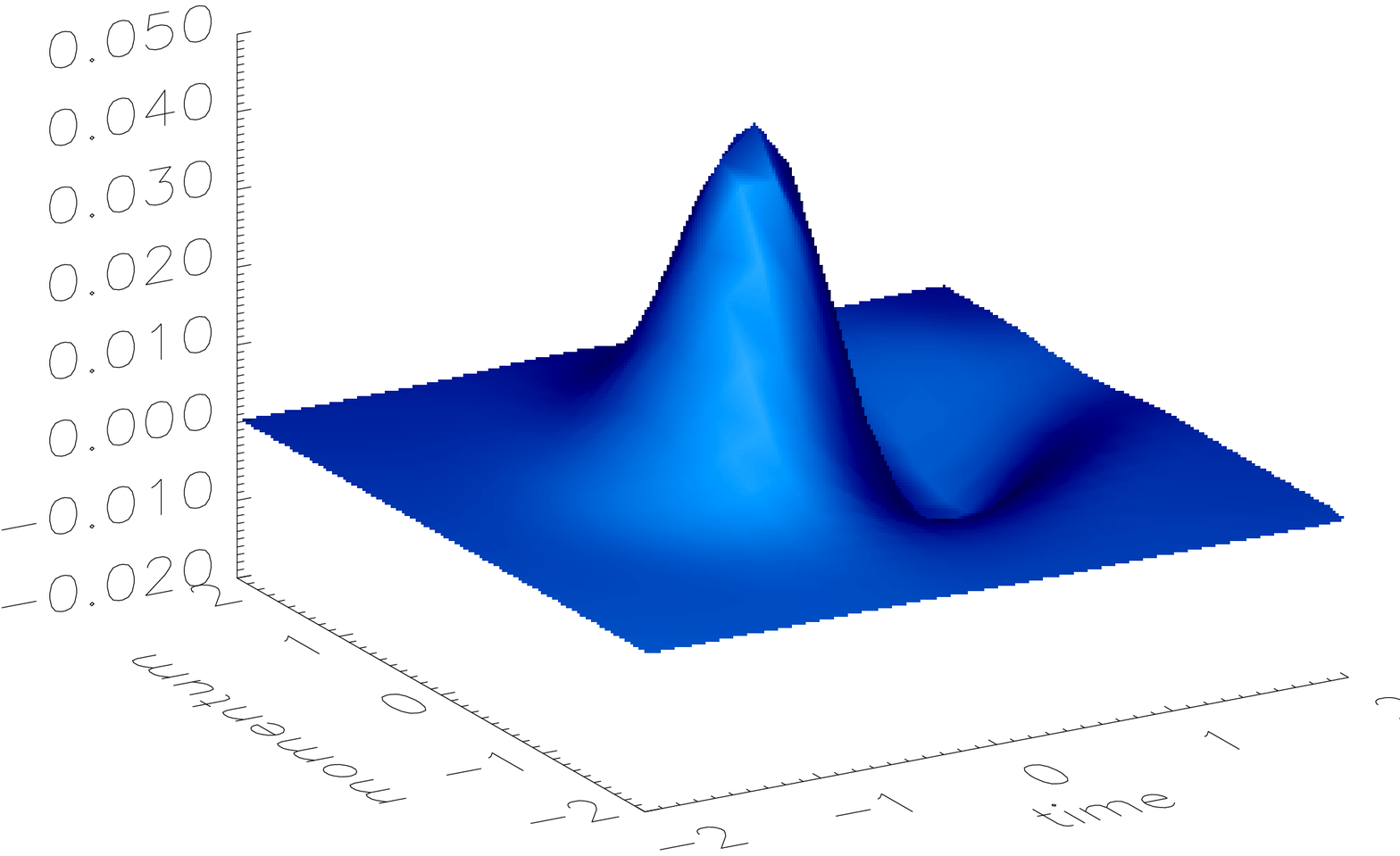,height=7.0cm,width=7.0cm,angle=0}}
\fcaption{The pair production rate, ${\tilde S}({\tilde p}_\parallel,{\tilde t})$, as a function of time and parallel momentum for a time dependent weak electric field charactarized by ${\tilde E}_0=0.5$, ${\tilde \sigma}= 1$ and $\tau = 0$ . All plotted values are dimensionless as described in the text.\label{fig4}}
\end{figure}  
 \begin{figure}[hbtp]
\centering{\epsfig{figure=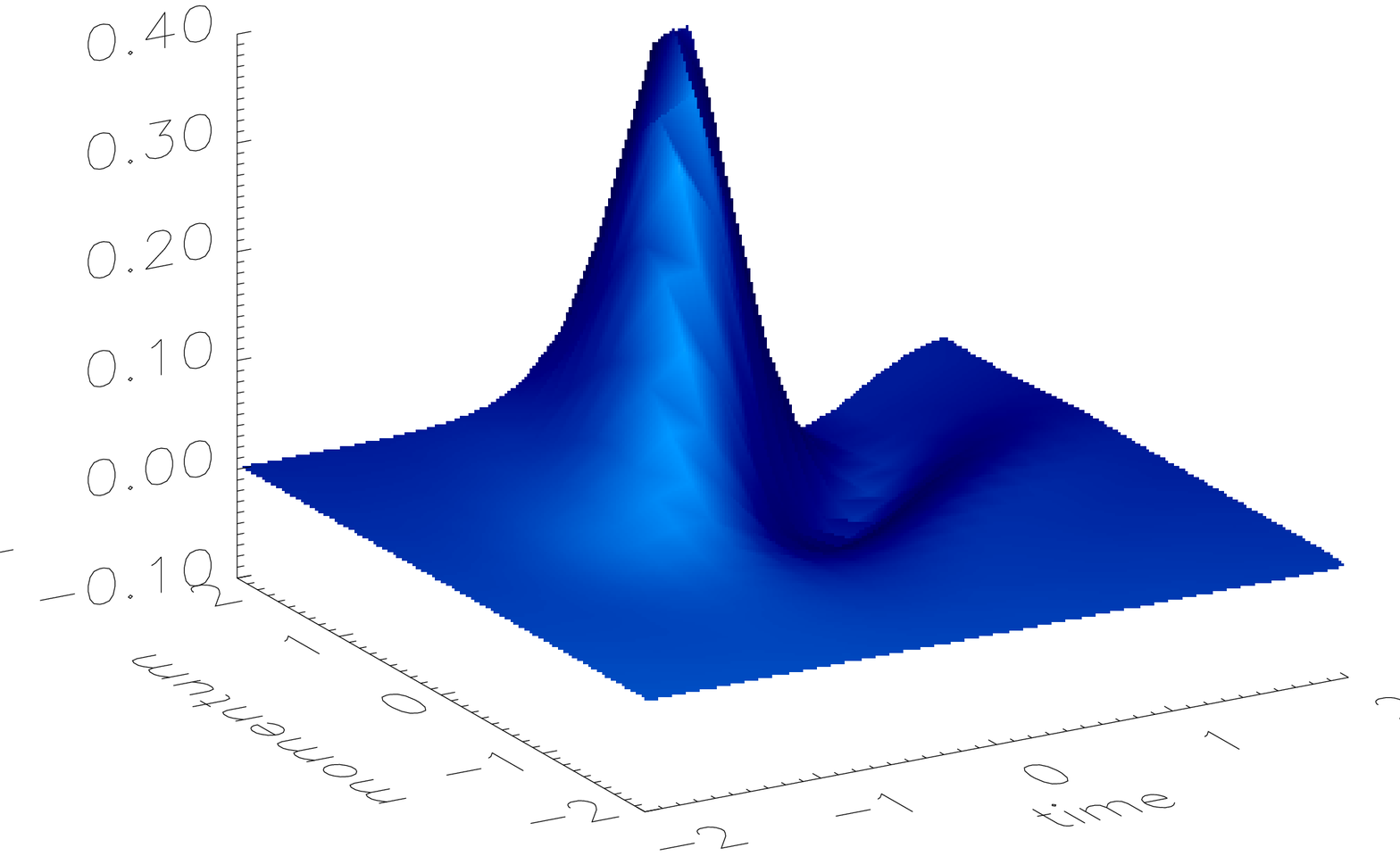,height=7.0cm,width=7.0cm,angle=0}}
\fcaption{The pair production rate, ${\tilde S}({\tilde p}_\parallel,{\tilde t})$, as a function of time and parallel momentum for a time dependent strong electric field charactarized by ${\tilde E}_0=1.5$, ${\tilde \sigma}= 1$ and $\tau = 0$ . All plotted values are dimensionless as described in the text.\label{fig5}}
\end{figure} 

The situation  changes if we allow the electric field to be time dependent.
We assume a Gaussian field at the dimensionsless time $\tau$ with the width ${\tilde \sigma} = \sigma\varepsilon_\perp$,
\be
{\tilde E}({\tilde t}) = {\tilde E}_0e^{-({\tilde t}-\tau)^2/{\tilde \sigma}^2}
\ee
and obtain
\be
{\tilde A}({\tilde t}) = -{\tilde E}_0\sigma\frac{\sqrt{\pi}}{2}\bigg[ {\rm Erf}[({\tilde t}-\tau)/{\tilde \sigma}] - {\rm Erf}[(-\tau-t_0)/{\tilde \sigma}]\bigg]\,.
\ee
The occuring error function is defined as
\be
{\rm Erf}(z) = \frac{2}{\sqrt{\pi}}\int_0^z e^{-x^2}dx \,.
\ee

Using this Ansatz for the field strength in Eq. (\ref{dimsource}) we obtain the numerical results plotted
in Figs. \ref{fig4} and \ref{fig5}. Therein all occuring values are dimensionsless. The electric field is non zero within a certain width ${\tilde \sigma}= 1$ for $\tau = 0$ around ${\tilde t} = 0$. Therefore the oscillations for times beyond the time where the electrical field is finite  are damped out. The pair production rate is peaked around small momenta for both weak and strong fields. For strong fields the distribution is shifted to positive momenta but remains still close to small parallel momenta. 
It is important to point out that the production of particles happens not only at rest ($p_\parallel=0$) what is assumed in many works also addressing  the back reaction problem, e.g.\cite{KES}. In contrary we find a non-trivial momentum dependence of the pair creation rate  depending on the field strength and on time.

%%%%%%%%%%%%%%%%%%%%%%%%%%%%%%%%%%%%%%%%%%%%%%%%%%%%%%%%%%%%%%%%%%%%%%%%%%%
\section{Summary}
%%%%%%%%%%%%%%%%%%%%%%%%%%%%%%%%%%%%%%%%%%%%%%%%%%%%%%%%%%%%%%%%%%%%%%%%
We have derived a quantum kinetic equation within a consistent field 
theoretical treatment which contains the creation of particle-antiparticle pairs in a time-dependent homogeneous electric field.  
For both fermions and bosons we obtain a source term providing a kinetic equation of  non-Markovian character. The source term is characterized by a pair production rate that contains a time  integration over the evolution of the distribution function and therefore involves memory effects.

 For the simple case of a constant electric field in low density limit and  Markovian approximation, we  analytically and numerically reproduce the results of earlier works\cite{Rau}. Since our approach is not restricted to constant fields, we have explored the dependence of the source term on a time dependent (model) electric field with a Gaussian shape. Within these two different Ans\"atze for the field, we have performed  investigations of the time structure of the source term. The production of pairs does not happen at rest only. We observe a non-trivial momentum dependence of the source term depending on the field strength and on time. 

The particle production source is dominated by two time scales: the memory time and the production time. The numerical results mainly show oscillations due to the dynamical phase and urge the need to include  the Maxwell equation to determine the electric field by physical boundary conditions (back reactions), work in this direction is in progress. 
Furthermore, it would be of great interest to extend this approach to the QCD case to explore the impact of a non-Markovian source term on the pre-equilibrium physics in ultrarelativistic heavy-ion collisions.

%%%%%%%%%%%%%%%%%%%%%%%%%%%%%%%%%%%%%%%%%%%%%%%%%%%%%%%%%%%%%%%%%%%% 
\nonumsection{Acknowledgment}
%%%%%%%%%%%%%%%%%%%%%%%%%%%%%%%%%%%%%%%%%%%%%%%%%%%%%%%%%%%%%%%%%%
The authors wish to thank J.M. Eisenberg,  V.G. Morozov and J. Rau for valuable discussions and comments. 
The authors
gratefully acknowledge the hospitality at the Tel Aviv University (S.S.), 
the University of Rostock
(S.A.S.), the JINR Dubna (D.B. and G.R.) and the GSI Darmstadt (V.D.T.)
where part of this work has been carried out.  This work was supported in
part by the Russian State Committee  of High-School Education under grant
No. 95-0-6/1-53, the Heisenberg-Landau program, the WTZ program of the BMBF
and by the HSP-III under the project No. 0037 6003.

\nonumsection{References}

\end{document}